\documentclass[12pt]{article}
\pdfoutput=1

\usepackage{amssymb,cite,color,comment,graphicx,hyperref,mathtools}
\usepackage[textheight=9in,textwidth=6.5in,letterpaper]{geometry}

\numberwithin{equation}{section}
\newcommand{\ab}[1]{\left|#1\right|}

\newcommand{\br}[1]{\left[#1\right]}
\newcommand{\cu}[1]{\left\{#1\right\}}
\newcommand{\pa}[1]{\left(#1\right)}

\newcommand{\ed}{\,\mathrm{d}}
\newcommand{\pd}{\,\partial}
\newcommand{\E}{\mathcal{E}}
\newcommand{\J}{\mathcal{J}}
\newcommand{\I}{\mathcal{I}}
\renewcommand{\L}{\mathcal{L}}

\newcommand{\R}{\mathbb{R}}
\newcommand{\SL}{\mathsf{SL}}
\newcommand{\U}{\mathsf{U}}

\begin{document}

\hfill

\begin{center}
	\vspace{2cm}
	{\LARGE{{Force-Free Electrodynamics\\\vspace{0.2cm} around Extreme Kerr Black Holes}}}

	\vspace{2cm}
	Alexandru Lupsasca, Maria J. Rodriguez and Andrew Strominger
	\vspace{2cm}

	{\it Center for the Fundamental Laws of Nature, Harvard University,\\
	Cambridge, MA 02138, USA}
	\vspace{1cm}
\end{center}

\begin{abstract}

\noindent Plasma-filled magnetospheres can extract energy from a spinning black hole and provide the power source for a variety of observed astrophysical phenomena. These magnetospheres are described by the highly nonlinear equations of force-free electrodynamics, or FFE. Typically these equations can only be solved numerically. In this paper we consider the FFE equations very near the horizon of a maximally spinning black hole, where the energy extraction takes place. Thanks to an enhanced conformal symmetry which appears in this near-horizon region, we are able to analytically obtain several infinite families of exact solutions of the full nonlinear equations.

\end{abstract}

\pagebreak

\tableofcontents

%%%%%%%%%%%%%%%%%%%% SECTION %%%%%%%%%%%%%%%%%%%%
\section{Introduction}
\label{sec:Introduction}

The sky contains a variety of objects, for example pulsars \cite{Bell:1968} and quasars \cite{Matthews:1977ds}, that produce extravagantly energetic signals such as collimated jets of electromagnetic radiation. In many cases, the energy source which powers these signals is suspected to be a rotating black hole surrounded by a magnetosphere with a plasma. Energy extraction from such a black hole is widely believed to be described by the highly nonlinear equations of force-free electromagnetism  \cite{Blandford:1977ds}. As our ability to observe these systems improves, a better quantitative and qualitative understanding of these interesting nonlinear equations is required. 

Most of the analyses of force-free electrodynamics have been numerical. A notable exception is the beautiful recent work of Brennan, Gralla and Jacobson \cite{Brennan:2013jla}, who found solutions by imposing the null condition $E^2=B^2$ (or, equivalently, $F^2=0$).  Other analytic work may be found in \cite{Curtis:1973,Blandford:1976,Blandford:1977ds,Menon:2007,Tanabe:2008wm,Brennan:2013ppa} and a review is \cite{Gralla:2014yja}. 

In this paper we continue the analytic approach in a different direction.  Energy extraction is possible only for rotating Kerr black holes, and the greater the rotation, the easier it becomes. Moreover it is a process that occurs near the black hole horizon, and is largely insensitive to the physics at spatial infinity. This suggests that much of the physics of force-free electrodynamic energy extraction can be captured by studying the near horizon region of maximally-rotating extreme Kerr black holes, such as the one in Cygnus X-1 \cite{Gou:2013dna}.

\begin{figure}[h!]
	\begin{center}
	\href{http://www.youtube.com/playlist?list=PLsrfyTK-g7cP-_8F7A5Zb71K_94_gaXgn}
	{\includegraphics[width=0.4\textwidth]{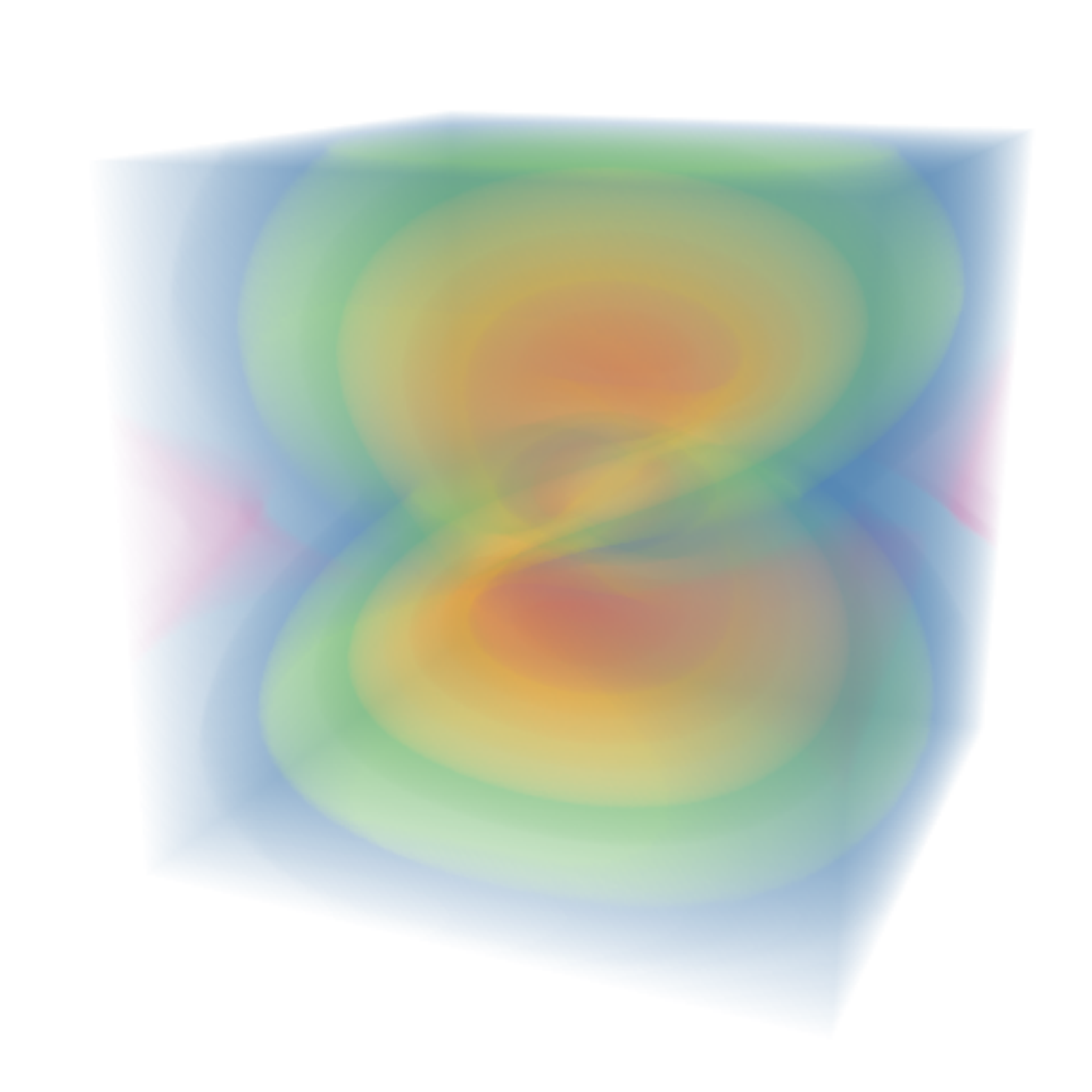}}\qquad
	\includegraphics[scale=0.7]{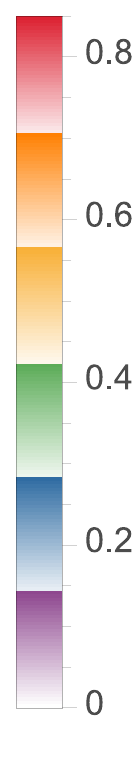}
	\href{http://www.youtube.com/playlist?list=PLsrfyTK-g7cP-_8F7A5Zb71K_94_gaXgn}
	{\includegraphics[width=0.4\textwidth]{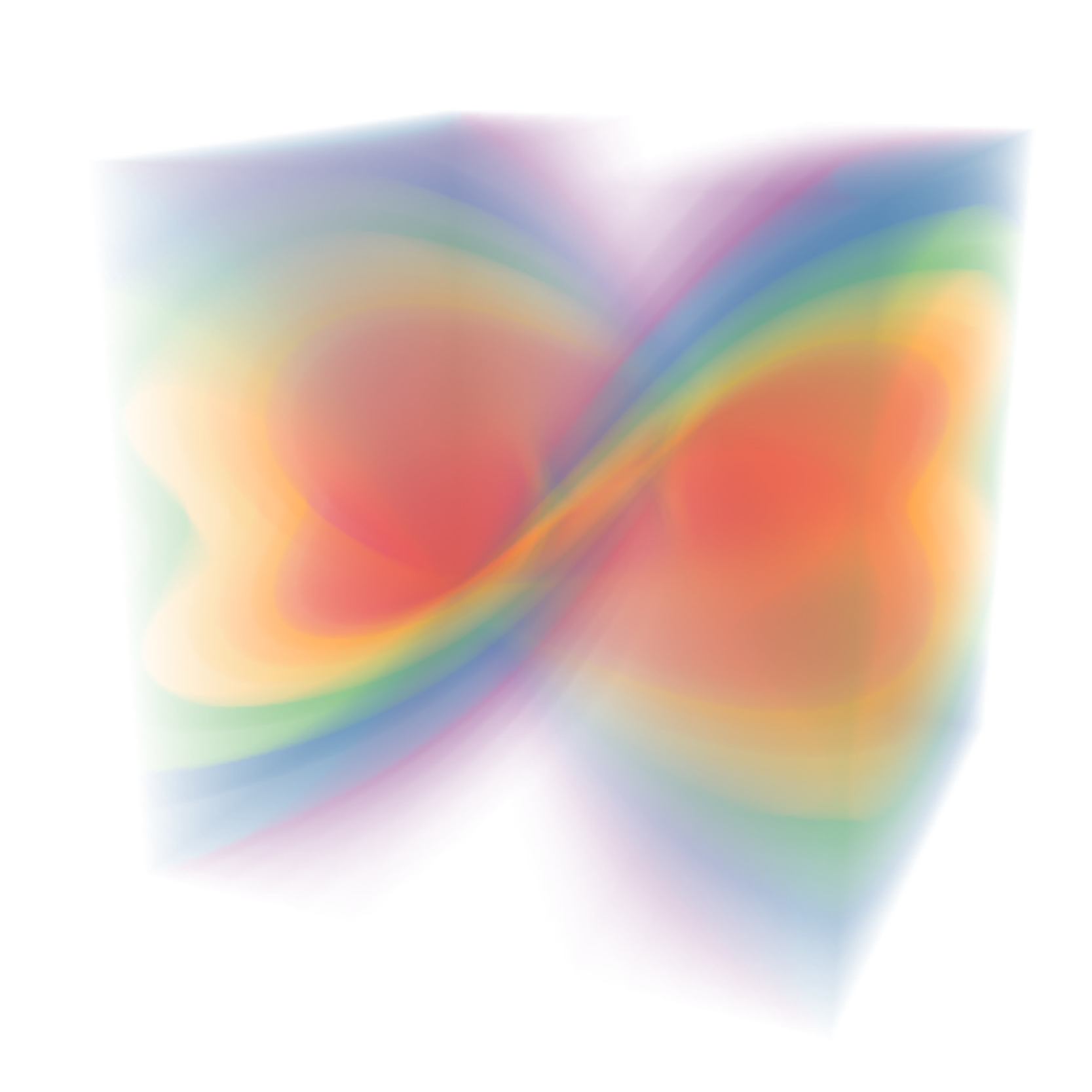}}
	\caption{Electric field strength $E^2$ (left) and magnetic field strength $B^2$ (right) evaluated at Poincar\'e time $t=0$ for the non-null solution $F=\Re F_{(2,0)}$. The black hole is the point at the center of the box. See subsection \ref{subsec:video} for further details, and \cite{video} for full animations.}
	\label{Fig:Intro}
	\end{center}
\end{figure}

Fortuitously, the dynamics of this region -- known as NHEK for Near Horizon Extreme Kerr -- is governed by an enhanced conformal symmetry which does not extend to the full Kerr geometry \cite{Bardeen:1999px,kcft}. This symmetry motivated the Kerr/CFT conjecture pertaining to the quantum structure of black holes \cite{Bredberg:2011hp} and also has potential consequences for observational astronomy \cite{Porfyriadis:2014fja,Hadar:2014dpa}. In this paper, it enables us to find large families of exact axisymmetric non-null ($F^2\neq0$) solutions of the equations of force-free electrodynamics, exhibiting a variety of complex behaviors. These axisymmetric solutions do not extract angular momentum, which precludes energy extraction. The more general case will be analyzed in a forthcoming paper \cite{inprogress}. One hopes that this analytic approach will enable a better understanding of astrophysical black hole magnetospheres and energy extraction. We illustrate some of the physical properties of these solutions in Figures \ref{Fig:Intro} and \ref{Fig:Current}.

The paper is organized as follows. We begin with a review of the force-free equations in section \ref{sec:FFE}, where we explain how they naturally arise in the context of electromagnetically-dominated systems. After a quick review of the NHEK geometry in subsections \ref{subsec:PoincareNHEK}--\ref{subsec:GlobalNHEK}, we express the force-free equations in the background of NHEK in a language propitious to the exploitation of spacetime symmetries. We conclude section \ref{sec:NHEK} with a prescription for computing, given a solution to these equations, the rate of energy and angular momentum extraction from both the event horizon of the black hole and the boundary of the throat.

Then in sections \ref{sec:Invariant} through \ref{sec:MariaSolution}, we present new solutions to the force-free equations in the background of NHEK. We obtain an $\SL(2,\R)\times\U(1)$-invariant solution (section \ref{sec:Invariant}), as well as an infinite tower of solutions, which lie in highest-weight representations of $\SL(2,\R)$ with arbitrary highest-weight $h\in\R-\cu{1}$ (subsection \ref{subsec:HWrep}). These all possess timelike currents ($\J^2<0$) and have $F^2\neq0$. Next, in subsection \ref{subsec:HWreph1} we turn to the null case $F^2=0$, where we obtain a highest-weight solution with highest weight $h=1$ and null current. Moreover, for each solution we also compute the corresponding fluxes of energy and angular momentum, both at the horizon of the black hole and at the boundary of the throat.

We also  expound upon a surprising property exhibited by our solutions, namely that their linear combinations are still solutions. Due to the non-linear character of the force-free equations, it is remarkable  that we can find new exact nonlinear solutions to these equations by forming linear superpositions, and we provide a simple  explanation of this phenomenon. 

Some mathematical details of our analysis have been relegated to Appendices \ref{appendix:Details} and \ref{appendix:Analysis}.

%%%%%%%%%%%%%%%%%%%% SECTION %%%%%%%%%%%%%%%%%%%%
\section{Force-free electrodynamics}
\label{sec:FFE}

This section contains a lightning review of force-free electrodynamics. Maxwell's equations are
\begin{align}
	\label{Maxwell}
	\nabla_\mu F^{\mu\nu}&=\J^\nu.
\end{align}
where $\J^{\nu}$ is the matter charge current and  $F_{\mu\nu}=\nabla_\mu A_\nu-\nabla_\nu A_\mu$. The electromagnetic stress-energy tensor is
\begin{align}
	\label{StressTensor}
	T_{\mathrm{EM}}^{\mu\nu}
	=F^{\mu\alpha}{F^\nu}_\alpha-\frac{1}{4}g^{\mu\nu}F_{\alpha\beta}F^{\alpha\beta}.
\end{align}
In general, this tensor is not covariantly conserved by itself. Indeed, Maxwell's equations imply that
\begin{align}
	\label{Lorentz}
	\nabla_\nu T_{\mathrm{EM}}^{\mu\nu}=-F_{\mu\nu}\J^\nu.
\end{align}
where the right hand side is the relativistic form of the Lorentz force density. The full stress-energy tensor
\begin{align}
	T^{\mu\nu}=T_{\mathrm{EM}}^{\mu\nu}+T_{\mathrm{matter}}^{\mu\nu} 
\end{align}
is always conserved
\begin{align}
	\label{Conservation}
	\nabla_\nu T^{\mu\nu}=0.
\end{align}
Force-free electrodynamics describes systems in which most of the energy resides in the electrodynamical sector of the theory, so that
\begin{align}
	\label{Assumption}
	T^{\mu\nu}\approx T_{\mathrm{EM}}^{\mu\nu}.
\end{align}
Under this assumption, the conservation of energy-momentum equation \eqref{Conservation} reduces to
\begin{align}
	\nabla_\nu T_{\mathrm{EM}}^{\mu\nu}=0.
\end{align}
This approximation is known as the ``force-free'' condition, since by \eqref{Lorentz} it is equivalent to the requirement that the Lorentz force density vanishes
\begin{align}
	\label{ForceFree}
	F_{\mu\nu}\J^\nu=0.
\end{align}
In the study of systems obeying this condition, the current $\J^\mu$ may be defined as the right hand side of $\nabla_\mu F^{\mu\nu}=\J^\nu$ rather than independently specified. A complete set of equations of motion for the electromagnetic sector is obtained by appending to Maxwell's equations the force free condition \eqref{ForceFree}.  In other words, a vector potential $A_\mu$ is a solution of force-free electrodynamics if and only if the resulting $F_{\mu\nu}$ and $\J_\mu$ obey \eqref{ForceFree}.\footnote{The initial data problem is subtle: see \cite{Komissarov:2002my} for a discussion.}

It is widely believed that astrophysical black holes are typically surrounded by magnetospheres composed of an electromagnetic plasma governed by these equations. Hence they are of both mathematical and physical interest. 

%%%%%%%%%%%%%%%%%%%% SECTION %%%%%%%%%%%%%%%%%%%%
\section{The NHEK geometry}
\label{sec:NHEK}

This section briefly reviews the geometry of Kerr and  the near-horizon NHEK region.  The force-free equations in NHEK and the conserved fluxes associated to the isometries are also described.

%%%%%%%%%%%%%%%%%%% SUBSECTION %%%%%%%%%%%%%%%%%%%
\subsection{Kerr geometry}
\label{subsec:Kerr}

The Kerr metric describes a rotating black hole with angular momentum $J$ and mass $M$. In Boyer-Lindquist coordinates $(\hat t,\hat r,\hat\theta,\hat\phi)$, its line element is
\begin{align}
	\label{Kerr}
	ds^2=&-\frac{\Delta}{\Sigma}\pa{\ed\hat t-a\sin^2{\hat\theta}\ed\hat\phi}^2
		+\frac{\Sigma}{\Delta}\ed\hat r^2+\frac{\sin^2\hat\theta}{\Sigma}\br{\pa{\hat r^2+a^2}\ed\hat\phi
		-a\ed\hat t}^2+\Sigma\ed\hat\theta^2,
\end{align}
where we set $c=G=1$ and defined
\begin{align}
	\Delta\equiv\hat r^2-2M\hat r+a^2,\qquad
	\Sigma\equiv\hat r^2+a^2\cos^2\hat\theta,\qquad
	a\equiv\frac{J}{M}.
\end{align}
There is an event horizon at
\begin{align}
	\label{OuterEventHorizon}
	\hat r_H=M+\sqrt{M^2-a^2},
\end{align}
from which it follows that the Kerr solution has a naked singularity unless $\ab{a}\le M$. This last bound is saturated by the so-called extreme Kerr solution, which carries the maximum allowed angular momentum
\begin{align}
	\ab{J}=M^2.
\end{align}

%%%%%%%%%%%%%%%%%%% SUBSECTION %%%%%%%%%%%%%%%%%%%
\subsection{The scaling limit}
\label{subsec:PoincareNHEK}

In this paper we are interested in the region very near the horizon of extreme Kerr, described by the so-called Near-Horizon Extreme Kerr (NHEK) geometry. It can be obtained by a near-horizon limiting procedure from the Kerr metric in usual Boyer-Lindquist coordinates \eqref{Kerr}. Following \cite{Bardeen:1999px}, define new dimensionless coordinates $(t,r,\theta,\phi)$ by 
\begin{align}
	\label{Zoom}
	t=\frac{\lambda\hat t}{2M},\qquad
	r=\frac{\hat r-M}{\lambda M},\qquad
	\theta=\hat\theta,\qquad
	\phi=\hat\phi-\frac{\hat t}{2M}.
\end{align}
In taking the limit $\lambda\to0$ while keeping these coordinates fixed, one is effectively ``zooming" into the region near the horizon. This procedure yields the NHEK line element in Poincar\'{e} coordinates
\begin{align}
	\label{eq:poincareNHEK}
	ds^2=2J\Gamma\br{-r^2\ed t^2+\frac{\ed r^2}{r^2}+\ed\theta^2+\Lambda^2(\ed\phi+r\ed t)^2},
\end{align}
where $t\in(-\infty,\infty)$, $r\in[0,\infty)$, $\theta\in[0,\pi]$, $\phi\sim\phi+2\pi$ and
\begin{align}
	\Gamma(\theta)\equiv\frac{1+\cos^2\theta}{2},\qquad
	\Lambda(\theta)\equiv\frac{2\sin\theta}{1+\cos^2\theta}.
\end{align}
The event horizon of the original extreme Kerr black hole is now located at
\begin{align}
	\label{zs}
	r_H=0.
\end{align}
In contrast with the original Kerr metric \eqref{Kerr}, the NHEK geometry is not asymptotically flat.

%%%%%%%%%%%%%%%%%%% SUBSECTION %%%%%%%%%%%%%%%%%%%
\subsection{Isometries}

A  crucial feature of the NHEK region is that the orignal  $\U(1)\times\U(1)$ Kerr isometry goup is enhanced to $\SL(2,\R)\times\U(1)$. This enhanced symmetry governs the dynamics of the near-horizon region of extreme Kerr. The $\U(1)$ rotational symmetry is generated by the Killing vector field
\begin{align}
	\label{xs}
	W_0=\pd_\phi.
\end{align}
The time translation symmetry becomes part of an enhanced $\SL(2,\R)$ isometry group generated by the Killing vector fields
\begin{align}
	\label{xss}
	H_0&=t\pd_t-r\pd_r,\\
	H_+&=\sqrt{2}\pd_{t},\\
	H_-&=\sqrt{2}\br{\frac{1}{2}\pa{t^2+\frac{1}{r^2}}\pd_t-tr\pd_r-\frac{1}{r}\pd_{\phi}}.
\end{align}
It is easily verified that these  satisfy the $\SL(2,\R)\times\U(1)$ commutation relations, namely:
\begin{align}
	\br{H_0,H_\pm}&=\mp H_\pm,\qquad\,\br{H_+,H_-}=2H_0,\\
	\br{W_0,H_\pm}&=0,\qquad\qquad\br{W_0,H_0}=0.
\end{align}
These symmetries do not leave the original Kerr horizon \eqref{zs} invariant and mix up the inside and outside of the original black hole.

%%%%%%%%%%%%%%%%%%% SUBSECTION %%%%%%%%%%%%%%%%%%%
\subsection{Global coordinates}
\label{subsec:GlobalNHEK}

The Poincar\'{e} coordinates $(t,r,\theta,\phi)$ cover only the part of the NHEK geometry outside the horizon of the original extreme Kerr.  Global coordinates $(\tau,\psi,\theta,\varphi)$ for NHEK are given by 
\begin{align}
	\label{eq:poincare2global}
	r=\frac{\cos{\tau}-\cos{\psi}}{\sin{\psi}},\qquad
	t=\frac{\sin{\tau}}{\cos{\tau}-\cos{\psi}},\qquad
	\phi=\varphi+\ln\ab{\frac{\cos{\tau}-\sin{\tau}\cot{\psi}}{1+\sin{\tau}\csc{\psi}}}.
\end{align}
In these new coordinates the line element \eqref{eq:poincareNHEK} becomes
\begin{align}
	\label{eq:globalNHEK}
	ds^2=2J\Gamma\br{\pa{-\ed\tau^2+\ed\psi^2}\csc^2{\psi}+\ed\theta^2
		+\Lambda^2\pa{\ed\varphi-\cot{\psi}\ed\tau}^2},
\end{align}
where $\tau\in(-\infty,\infty)$, $\psi,\theta\in[0,\pi]$ and $\varphi\sim\varphi+2\pi$. In global coordinates, a useful complex basis for the $\SL(2,\R)\times\U(1)$ Killing vectors is:
\begin{align}
	L_\pm&=ie^{\pm i\tau}\sin{\psi}\pa{-\cot{\psi}\pd_\tau\mp i\pd_\psi+\pd_\varphi},\\
	L_0&=i\pd_\tau,\\
	Q_0&=-i\pd_\varphi.
\end{align}
These obey
\begin{align}
	\br{L_0,L_\pm}&=\mp L_\pm,\qquad\,\br{L_+,L_-}=2L_0,\\
	\br{Q_0,L_\pm}&=0,\qquad\qquad\br{Q_0,L_0}=0,
\end{align}
and are related to \eqref{xs} and \eqref{xss} by
\begin{align}
	Q_0=-iW_0,\qquad
	L_0=\frac{i}{\sqrt{2}}\pa{\frac{1}{2}H_++H_-},\qquad
	L_{\pm}=\mp H_0+\frac{i}{\sqrt{2}}\pa{\frac{1}{2}H_+-H_-}.
\end{align}
The inverse relation is
\begin{align}
	\label{eq:H2L}
	H_0=\frac{L_--L_+}{2},\qquad
	H_+=-\frac{i}{\sqrt{2}}\pa{2L_0+L_++L_-},\qquad
	H_-=-\frac{i}{2\sqrt{2}}\pa{2L_0-L_+-L_-}.
\end{align}

%%%%%%%%%%%%%%%%%%% SUBSECTION %%%%%%%%%%%%%%%%%%%
\subsection{Force-free electrodynamics in NHEK}
\label{subsec:FFEinNHEK}

We now turn to the study of force-free electrodynamics in the NHEK geometry. It is convenient to use  differential form notation (see e.g. \cite{Carroll:1997ar} for conventions) in which  $F\equiv\ed A$ denotes the electromagnetic field strength, and the complete force-free equations of motion  for $F$ are 
\begin{align}
	\label{forcefree1}
	\ed F&=0,\\
	\label{forcefree2}
	\ed^\dagger F&=\J,\\
	\label{forcefree3}
	\J\wedge\star F&=0,
\end{align}
with $\star$ the Hodge dual,  $\wedge$ the wedge product and $\ed^\dagger$ the adjoint of the exterior derivative $d$.

In general these equations are highly nonlinear and can only be solved numerically. However in NHEK the symmetries can be exploited to simplify the analysis. Given one solution of the force-free equations, another can always be generated by the action of an isometry. Therefore the solutions must lie in representations of $\SL(2,\R)\times\U(1)$, which are studied for example in \cite{vilenkin}. In this paper, we look for axisymmetric solutions which lie in the so-called highest-weight representations of $\SL(2,\R)$. These contain an element obeying the relations
\begin{align}
	\label{highest1}
	\L_{L_+}F&=0,\\
	\label{highest2}
	\L_{L_0}F&=hF,\\
	\label{highest3}
	\L_{Q_0}F&=0,
\end{align}
where $\L_V$ denotes the Lie derivative with respect to the vector field $V$ and $h$ is a constant characterizing  the representation. The last condition requires that $F$ be $\U(1)$-invariant, while the first two conditions state that $F$ is in a highest-weight representation of $\SL(2,\R)$ with weight $h$. 

In the ensuing analysis we will find  force-free solutions obeying \eqref{highest1}--\eqref{highest3} for every real value of $h$. From each of these, an infinite family is obtained by the action of $\SL(2,\R)$. Since $L_+$ is complex, all of these solutions are complex. However we will show that the real and imaginary parts of these solutions surprisingly also solve the force-free equations and hence provide physical field configurations.  

%%%%%%%%%%%%%%%%%% SUBSECTION %%%%%%%%%%%%%%%%%%%%
\subsection{Energy and angular momentum flux}
\label{subsec:Fluxes}

The NHEK geometry possesses an axial $\U(1)$ symmetry generated by $W_0=\pd_\phi$, as well as a time-translation symmetry generated by $H_+=\sqrt{2}\pd_t$. It is therefore natural to define energy and angular momentum in NHEK as the conserved quantities associated with $H_+$ and $W_0$, respectively\footnote{This definition coincides with that of angular momentum in the Kerr spacetime \eqref{Kerr} because under the ``zooming'' procedure \eqref{Zoom}, the axial symmetry generator $\pd_{\hat\phi}$ of the Kerr black hole descends precisely to the NHEK generator $W_0=\pd_{\phi}$. However, the notions of energy in NHEK and Kerr differ because the generator $\pd_{\hat t}$ of the Kerr metric does not simply correspond to the NHEK energy generator $H_+=\sqrt{2}\pd_t$ -- instead, it becomes mixed with the angular momentum.}.

Given a solution to the force-free equations \eqref{forcefree1}--\eqref{forcefree3}, one may compute its stress-energy-momentum tensor $T_{\mu\nu}=T^\mathrm{EM}_{\mu\nu}$ and thence obtain the associated NHEK energy current $\I^E$ and angular momentum current $\I^L$
\begin{align}
	\label{Currents}
	\I_\nu^E&\equiv H_+^\mu T_{\mu\nu},\\
	\I_\nu^L&\equiv W_0^\mu T_{\mu\nu}.
\end{align}
By \eqref{Conservation} and the Killing equation, these currents are conserved:
\begin{align}
	\nabla^\nu\I_\nu^{E,L}=0.
\end{align}
Therefore, integrating either one of them over any region $R$ in the bulk yields
\begin{align}
	\int_R\!\ed^4x\sqrt{-g}\,\nabla^\nu\I_\nu^{E,L}=0.
\end{align}

\begin{figure}[h!]
	\begin{center}
	\includegraphics[width=0.4\textwidth,height=0.7\textwidth]{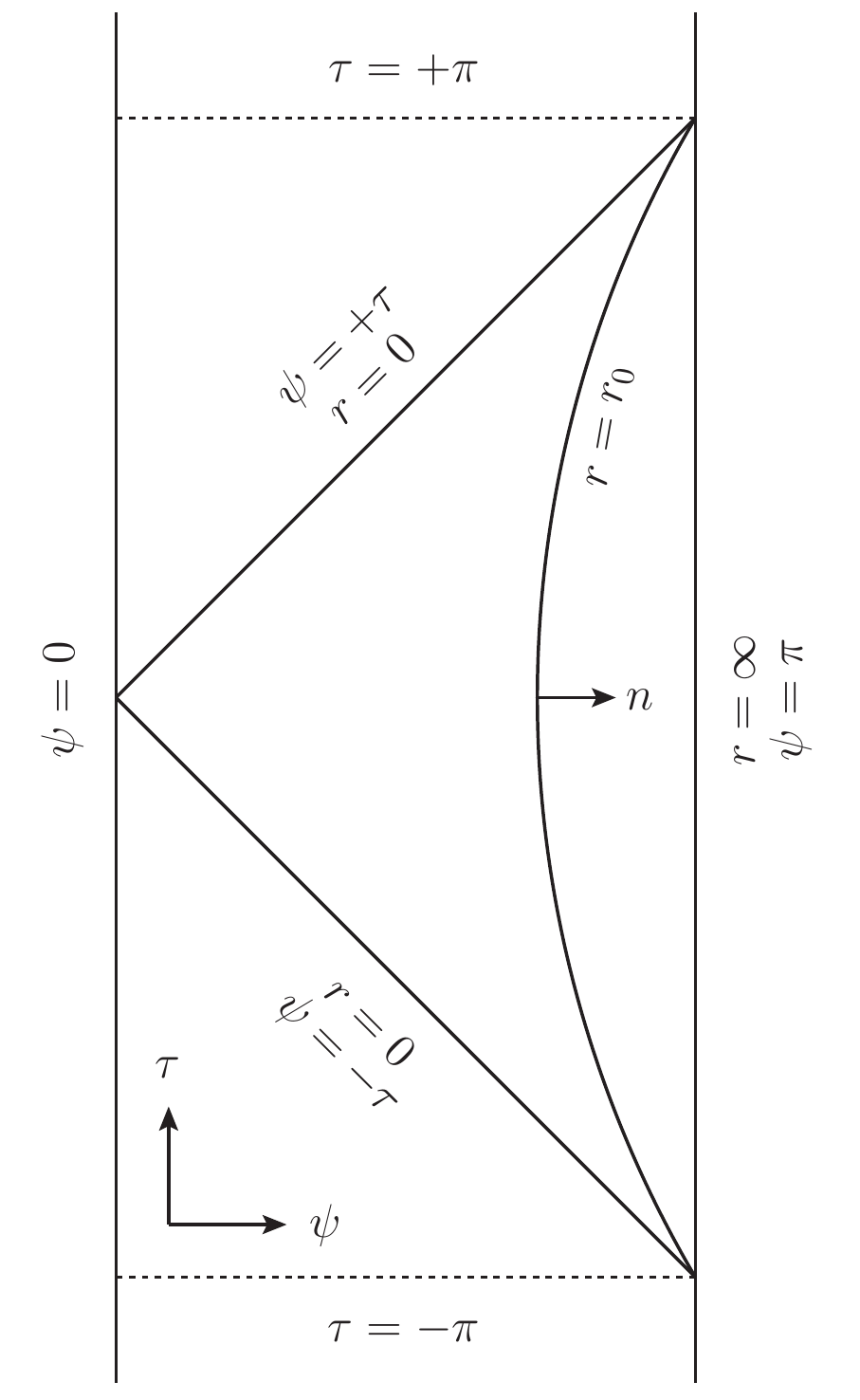}
	\caption{Penrose diagram for the NHEK geometry (\ref{eq:poincareNHEK}). The horizon is at $r=r_H=0$.}
	\label{Fig:Penrose}
	\end{center}
\end{figure}

Now suppose that $R$ is the entirety of the NHEK Poincar\'{e} patch, as depicted in Figure \ref{Fig:Penrose}. Then by Stokes' Theorem, the previous equation implies the energy conservation relation
\begin{align}
	\label{eq:fluxcondition}
	\Delta E_H^++\Delta E_H^-+\Delta E_B=0,
\end{align}
where $\Delta E_H^+$ denotes the total energy crossing into the future horizon ($\psi=+\tau$), $\Delta E_H^-$ is minus the energy coming out of the past horizon ($\psi=-\tau$), and $\Delta E_B$ is the total energy extracted from the boundary of the throat (at $r\to\infty$ in Poincar\'{e} coordinates or $\psi=\pi$ in global coordinates). These quantities are most conveniently computed in global coordinates, which are smooth across the horizon, as
\begin{align}
	\label{eq:totalfluxes}
	\Delta E_H^+&=\int_0^\pi\!\ed\tau\int_0^\pi\!\ed\theta\int_0^{2\pi}\!\ed\varphi\ \E_H^+,\\
	\Delta E_H^-&=\int_{-\pi}^0\!\ed\tau\int_0^\pi\!\ed\theta\int_0^{2\pi}\!\ed\varphi\ \E_H^-,\\
	\Delta E_B&=\int_{-\pi}^\pi\!\ed\tau\int_0^\pi\!\ed\theta\int_0^{2\pi}\!\ed\varphi\ \E_\infty,
\end{align}
where the integrands correspond to the energy flux density per solid angle on the horizon and the boundary of the throat
\begin{align}
	\label{eq:fluxes}
	\E_H^\pm&\equiv\sqrt{\gamma}\pa{\pm\sqrt{2}H_+^\nu}\I^E_\nu\Big|_{\psi=\pm\tau},\\
	\E_\infty&\equiv\sqrt{-\sigma}\,n^\nu\I^E_\nu\Big|_{\psi=\pi}.
\end{align}
In these expressions, $\sigma$ is the induced 3-metric on the boundary of the throat and $n$ is the outward unit vector normal to this boundary, while $\gamma$ denotes the 2-metric on the event horizon, which has null generator $H_+$ (see \cite{Lasota:2013kia} and Appendix \ref{appendix:Details} for details). A completely analogous story holds for the angular momentum flux, with $\L$ and $L$ replacing $\E$ and $E$, respectively.

In the following sections, we will evaluate the energy and angular momentum densities at the horizon $r=r_H$ ($\E_H$ and $\L_H$) and at the boundary $r\to\infty$ ($\E_{\infty}$ and $\L_{\infty}$) of NHEK to show that our force-free solutions do indeed produce non-trivial fluxes. Some of the details involved in these calculations are presented in Appendix \ref{appendix:Details}.

%%%%%%%%%%%%%%%%%%%% SECTION %%%%%%%%%%%%%%%%%%%%
\section{Maximally symmetric  solution }
\label{sec:Invariant}

In this section we construct the unique solution with the full $\SL(2,\R)\times\U(1)$ symmetry.

Consider the vector potential
\begin{align}
	A_{(0,0)}&\equiv P_0\pa{\cot{\psi}\ed\tau-i\ed\psi}
	\label{typeIIE}\\
	&=-\frac{iP_0}{2J\Gamma}\pa{\Phi\,L_++Q_0},
	\label{typeIIEbis}
\end{align}
where $P_0$ is a function of $\theta$ only and 
\begin{align}
	\label{Phi}
	\Phi(\tau,\psi)\equiv e^{-i\tau}\sin{\psi}.
\end{align}
For the maximally symmetric case we could actually eliminate the $\Phi L_+$ term  in \eqref{typeIIEbis} by a gauge transformation\footnote{An equivalent gauge potential is
\begin{align}
	\tilde{A}_{(0,0)}\equiv\frac{iQ_0}{Q_0\cdot Q_0}
	=-2J\Gamma\csc^2{\psi}\br{\pa{1-\Lambda^2}\cot{\psi}\ed\tau+\Lambda^2\ed\varphi}.
\end{align}}: we keep it to facilitate the generalizations of the next section. $A_{(0,0)}$ is $\SL(2,\R)\times\U(1)$-invariant
\begin{align}
	\L_{L_\pm}A_{(0,0)}=\L_{L_0}A_{(0,0)}=\L_{Q_0}A_{(0,0)}=0.
\end{align}
The field strength is 
\begin{align}
	F_{(0,0)}&=-P_0\csc^2{\psi}\ed\tau\wedge\ed\psi
	+P_0'\pa{\cot{\psi}\ed\tau\wedge\ed\theta-i\ed\psi\wedge\ed\theta}\\
	&=-\frac{i}{(2J\Gamma)^2}\br{
	P_0\pa{\Phi\,L_+\wedge L_0-\Phi\cot{\psi}\,L_+\wedge Q_0-L_0\wedge Q_0}
		-P_0'\pa{\Phi\,L_+\wedge\Theta+Q_0\wedge\Theta}},
\end{align}
where we have defined a 1-form
\begin{align}
	\Theta\equiv2J\Gamma \ed\theta.
\end{align}
Here and hereafter, it is understood that in this paper we use the same symbol (e.g. `$L_+$') to denote both a vector field and its associated 1-form, and rely on the context to distinguish between the two uses.

The Hodge dual of $F_{(0,0)}$   is
\begin{align}
	\star F_{(0,0)}=\frac{i}{(2J\Gamma)^2\Lambda}\br{P_0\,Q_0\wedge\Theta
		-P_0'\pa{\Phi\cot{\psi}\,L_+\wedge Q_0+L_0\wedge Q_0}}.
\end{align}
The current $\J_{(0,0)}\equiv\ed^\dagger F_{(0,0)}$ will evidently obey the force free condition if it is proportional to $Q_0$. This requires \begin{align}
	\label{InvariantTheta}
	\pd_\theta\pa{\Lambda\pd_\theta P_0}=0,
\end{align}
whose general solution is 
\begin{align}
	P_0(\theta)=C_0+D_0\int\frac{\ed\theta}{\Lambda(\theta)},
\end{align}
for some constants $C_0$ and $D_0$. In order for $P_0$ to be nonsingular on $\br{0,\pi}$ and $F$ real, one must set $D_0=0$ and $C_0$ real (recall that $Q_0$ is imaginary). One then finds 
\begin{align}
	\J_{(0,0)}=-\frac{iC_0}{(2J\Gamma)^2}Q_0,\qquad
	F_{(0,0)}=-C_0\csc^2{\psi}\ed\tau\wedge\ed\psi,
\end{align}
and
\begin{align}
	\J_{(0,0)}\wedge\star F_{(0,0)}=0. 
\end{align}
Hence we have a solution to \eqref{forcefree1}--\eqref{highest3} with $h=0$. Note that for this solution
\begin{align}
	F_{(0,0)}^2=-\frac{2C_0^2}{(2J\Gamma)^2}
\end{align}
is negative which indicates the field is largely electric.

%%%%%%%%%%%%%%%%%%% SUBSECTION %%%%%%%%%%%%%%%%%%%
\section{Axisymmetric highest-weight representations: generic case}
\label{sec:Solutions}

In this section we construct  large families of $U(1)$ axisymmetric solutions to the force-free equations in highest-weight representations labelled by a real parameter $h$. The solutions degenerate for the case $h=1$. A separate treatment  of this case is given in the next section. 
%%%%%%%%%%%%%%%%%%% SUBSECTION %%%%%%%%%%%%%%%%%%%

\subsection{Highest-weight solutions }
\label{subsec:HWrep}

An axisymmetric highest weight vector potential with weight $h$ obeys 
\begin{align}
	\L_{L_+}A_{(h,0)}&=0,\\
	\L_{L_0}A_{(h,0)}&=h\,A_{(h,0)},\\
	\L_{Q_0}A_{(h,0)}&=0.
\end{align}
These conditions are solved by
\begin{align}
	A_{(h,0)}&\equiv\Phi^h P_h\pa{\cot{\psi}\ed\tau-i\ed\psi}
	\label{typeIIA}\\
	&=-\frac{i\Phi^h P_h}{2J\Gamma}\pa{\Phi\,L_++Q_0},
\end{align}
where $P_h$ is a function of the $\theta$. $\Phi(\tau,\psi)$ was introduced in \eqref{Phi} and obeys 
\begin{align}
	\label{ScalarHighest}
	\L_{L_+}\Phi^h=0,\qquad
	\L_{L_0}\Phi^h=h\Phi^h,\qquad
	\L_{Q_0}\Phi^h=0.
\end{align}
For $h=0$ this vector potential reduces to the $\SL(2,\R)\times\U(1)$-invariant potential $A_{(0,0)}$ analyzed in the previous section. The  field strength $F_{(h,0)}\equiv\ed A_{(h,0)}$ is given by 
\begin{align}
	F_{(h,0)}&=-\Phi^h\br{(h-1)P_h\csc^2{\psi}\ed\tau\wedge\ed\psi
		+P_h'\pa{\cot{\psi}\ed\tau\wedge\ed\theta-i\ed\psi\wedge\ed\theta}}\\
	&=\frac{i\Phi^h}{(2J\Gamma)^2}\cu{
		(h-1)P_h\br{\pa{\Phi\,L_++Q_0}\wedge L_0-\Phi\cot{\psi}\,L_+\wedge Q_0}
		+P_h'\pa{\Phi\,L_++Q_0}\wedge \Theta}.
\end{align}
The Hodge dual of this expression is
\begin{align}
	\label{HodgePrimary}
	\star F_{(h,0)}=-\frac{i\Phi^h}{(2J\Gamma)^2\Lambda}\br{
		(h-1)P_h\,Q_0\wedge\Theta
		+P_h'\pa{\Phi\cot{\psi}\,L_+\wedge Q_0+L_0\wedge Q_0}}.
\end{align}
If the function $P_h$ satisfies 
\begin{align}
	\label{thetaode}
	\pd_\theta\pa{\Lambda\pd_\theta P_h}+h(h-1)\Lambda P_h=0,
\end{align}
then 
\begin{align}
	\label{CurrentPrimary}
	\J_{(h,0)}=\frac{i\Phi^hP_h}{(2J\Gamma)^2}(h-1)Q_0,
\end{align}
is proportional to $Q_0$, guaranteeing satisfaction of the force free condition $\J_{(h,0)}\wedge\star F_{(h,0)}=0$. Observe that when $h=1$,  $\J_{(1,0)}$ vanishes and we obtain a solution to the free Maxwell equations which trivially solves the force-free equations.

\begin{figure}[h!]
	\begin{center}
	\includegraphics[width=0.75\textwidth]{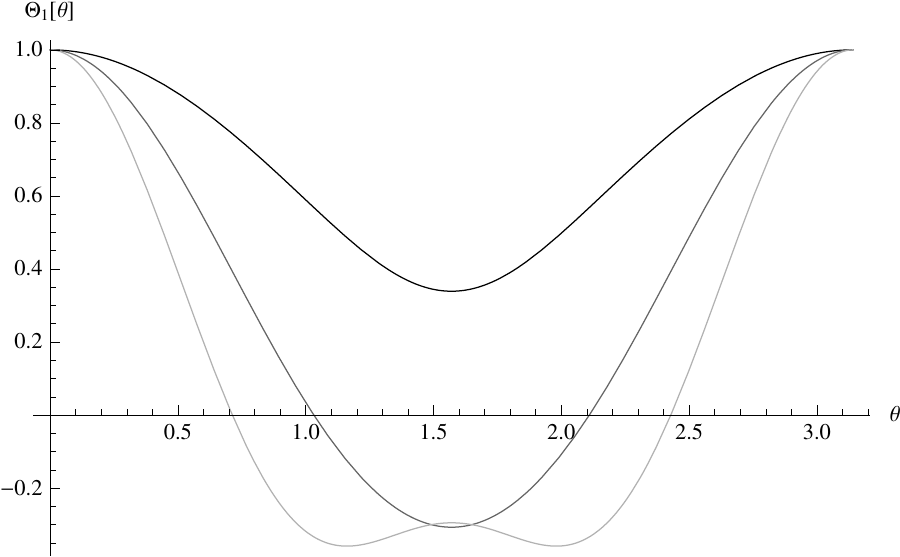}
	\caption{The figure shows $P_h(\theta)$ with $h=2,3,4$ (respectively {\it black} to {\it lighter grays}).}
	\label{Fig:Theta}
	\end{center}
\end{figure}

The differential equation \eqref{thetaode}, which defines a generalized Heun's function, is analyzed in Appendix \ref{appendix:Analysis}. It has a unique nonsingular solution up to a multiplicative constant. There is  no closed form expression but it may be expanded as
\begin{align}
	P_h(\theta)=\sum_{n=0}^\infty a_n\sin^{2n}{\theta},
\end{align}
where
\begin{align}
	a_{n+1}=B_na_n+C_na_{n-1},
\end{align}
and
\begin{align}
	B_n=\frac{6n^2-h(h-1)}{4(n+1)^2},\qquad
	C_n=-\frac{(2n-h-2)(2n+h-3)}{8(n+1)^2}.
\end{align}
This power series converges everywhere on the domain of interest $\theta\in[0,\pi]$. Moreover, it renders manifest the reflection symmetry of $P_h$ about the $\theta=\pi/2$ plane. In Figure \ref{Fig:Theta} we illustrate $P_h$ for representative values of $h$.

We note that  $F^2$ is in general nonzero and complex:
\begin{align}
	\label{Fsquaredcomplex}
	F_{(h,0)}^2=-2\pa{\frac{\Phi^h}{2J\Gamma}}^2\br{(h-1)^2P_h^2+P_h'^2}.
\end{align}

%%%%%%%%%%%%%%%%%%% SUBSECTION %%%%%%%%%%%%%%%%%%%
\subsection{Descendants }
\label{subsec:Descendants}

$\SL(2,\R)$ invariance of NHEK guarantees that any finite $\SL(2,\R)$ transformation of the above highest weight solutions are also solutions. If the equations were linear, this would immediately imply that the $\SL(2,\R)$ descendants (i.e. the fields obtained by acting with the raising operator $\L_{L_-}$) of these solutions, which are infinitesimal transformations,  are also solutions. Despite the nonlinearity of the equations, the descendants also turn out to solve the force-free equations.  

The reason for this is simple. If we start with the vector potential $A_{(h,k)}\equiv\L_{L_-}^kA_{(h,0)}$ given by the $k^\text{th}$ descendant, the resulting dual field strength and current  $\star F_{(h,k)}$ and $\J_{(h,k)}$ will also be $k^\text{th}$ descendants. Since both the highest weight dual field strength and current are proportional to $Q_0$ and $\L_{L_-}Q_0=0$, the descendants are all also proportional to $Q_0$. This guarantees that $\star F_{(h,k)}\wedge \J_{(h,k)}=0$ and the force-free equations are satisfied. 

To be explicit the $k^\text{th}$ descendant
\begin{align}
	\label{VectorDescendants}
	A_{(h,k)}&\equiv\L_{L_-}^kA_{(h,0)}
	=-\frac{iP_h}{2J\Gamma}\sum_{n=0}^k\binom{k}{n}\Phi_{(h,k-n)}\L_{L_-}^n\pa{\Phi\,L_++Q_0},
\end{align}
with 
\begin{align}
	\Phi_{(h,k)}&\equiv\L_{L_-}^k\Phi^h\\
	&=-\frac{2\Gamma\pa{h+\frac{1}{2}}\Gamma\pa{2h+k}}{\sqrt{\pi}\Gamma(2h)(k+1)}\Phi^{k+h}
	\sum_{n=0}^k\frac{(-1)^{k+n}\Gamma\pa{h+\frac{1}{2}}}{\Gamma\pa{h+n-\frac{1}{2}}}
	\binom{k+1}{k+2-2n}\pa{\cot{\psi}}^{k+2-2n}
\end{align}
yields the currents
\begin{align}
	\label{CurrentDescendants}
	\J_{(h,k)}=\L_{L_-}^k\J_{(h,0)}
	=\frac{i\Phi_{(h,k)}P_h}{(2J\Gamma)^2}(h-1)Q_0
\end{align}
and solves  the force-free equations \eqref{forcefree1}--\eqref{forcefree3}.

%%%%%%%%%%%%%%%%%%% SUBSECTION %%%%%%%%%%%%%%%%%%%
\subsection{Reality conditions}
\label{subsec:Physical}

So far the solutions of this section have been complex.  Physically we are interested in real solutions.  In general the real or imaginary part of a solution to a nonlinear equation will not itself solve the equation. However in the present case, taking the real part of the vector potential leads to dual field strengths and currents which are the real parts of the original ones. Since $Q_0$ has constant phase, the real  or imaginary parts of anything proportional to $Q_0$ is itself proportional to $Q_0$.  It follows that the real  or imaginary parts of all the solutions,  $\Re A_{(h,k)}$ and $\Im A_{(h,k)}$, are themselves solutions, although no longer simple descendants of a highest-weight solution.

It is important to note that these physical solutions no longer have a complex $F^2$ as in \eqref{Fsquaredcomplex}. Rather, we find that $F^2$ may be positive or negative at different points in the spacetime.

%%%%%%%%%%%%%%%%%%% SUBSECTION %%%%%%%%%%%%%%%%%%%
\subsection{A very general solution}
\label{subsec:video}

The arguments of the preceding two subsections are readily generalized to imply that the general linear combination
\begin{align}
	\label{gen}
	A(c,d)=\int_{-\infty}^\infty\!\ed h\,\sum_{k=0}^\infty\br{c_k(h)\Re A_{(h,k)}+d_k(h)\Im A_{(h,k)}}
\end{align}
for arbitrary real functions $c_k(h)$ and $d_k(h) $ is a real solution to the force free equations.  This follows because every term on the right hand side of \eqref{gen} gives both a $\star F$ and a $\J$ proportional to $Q_0$. Hence the force-free equation $\star F(c,d)\wedge \J(c,d)=0$ is satisfied. 

What has happened here is that we have effectively linearized the equations: the conditions that $\star F$ and $\J$ be proportional to $Q_0$ are linear conditions which imply the full nonlinear equation. Solutions of linear equations can always be added, hence the general solution \eqref{gen}.

\begin{figure}[h!]
	\begin{center}
	\href{http://www.youtube.com/playlist?list=PLsrfyTK-g7cP-_8F7A5Zb71K_94_gaXgn}
	{\includegraphics[width=0.4\textwidth]{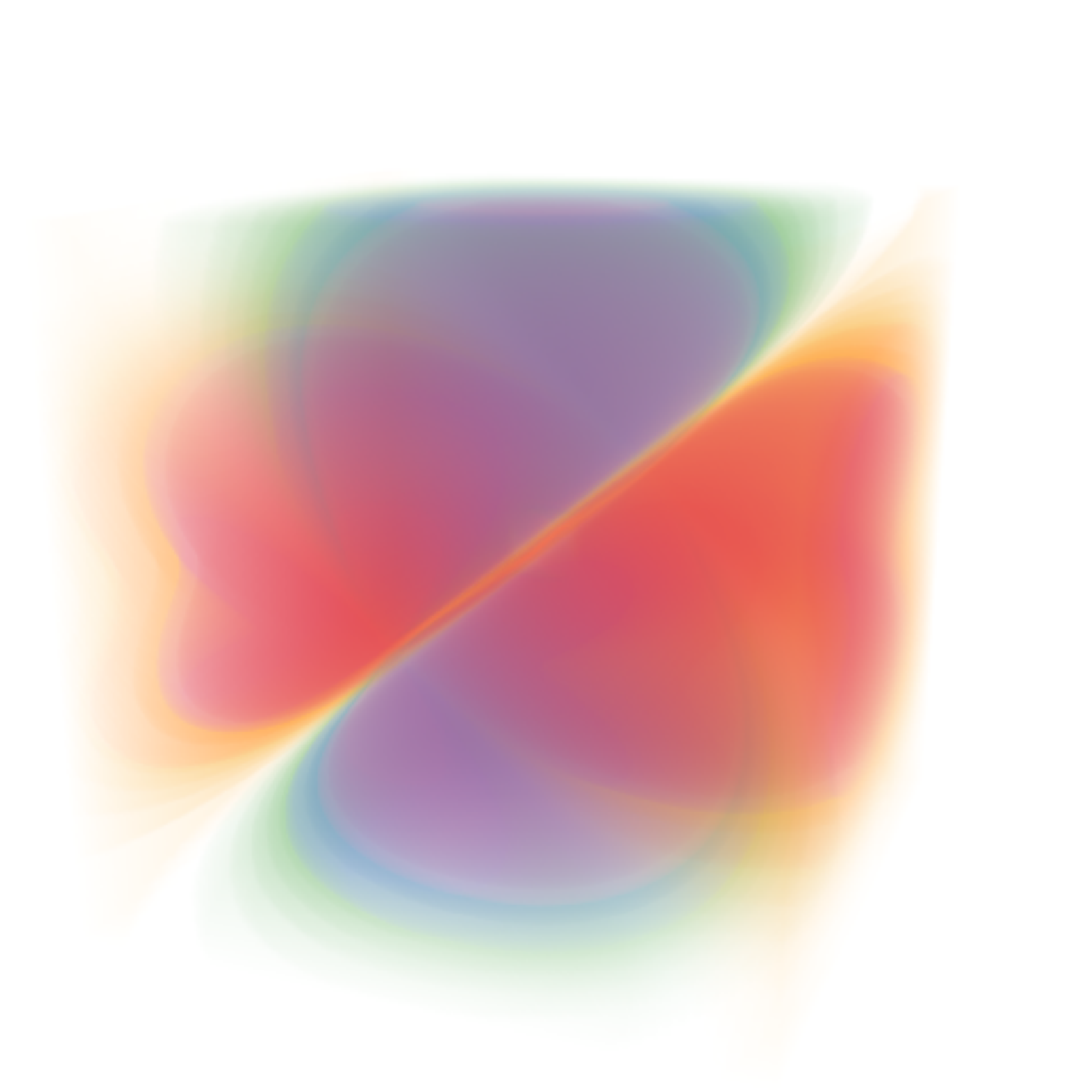}}
	\includegraphics[scale=0.7]{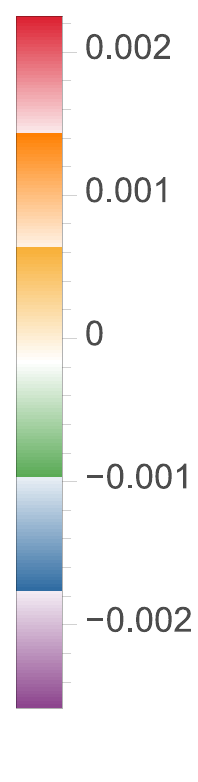}
	\href{http://www.youtube.com/playlist?list=PLsrfyTK-g7cP-_8F7A5Zb71K_94_gaXgn}
	{\includegraphics[width=0.4\textwidth]{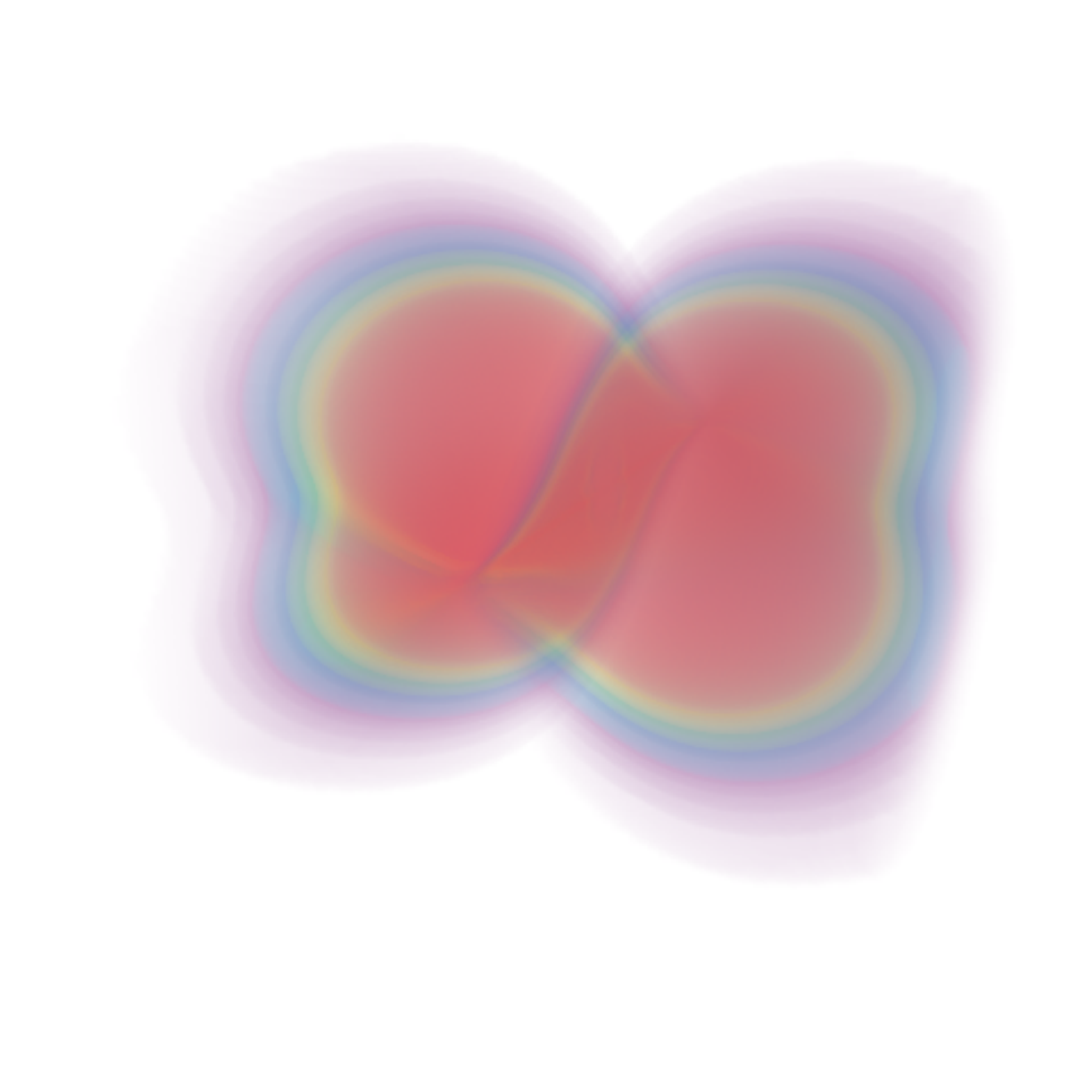}}
	\includegraphics[scale=0.7]{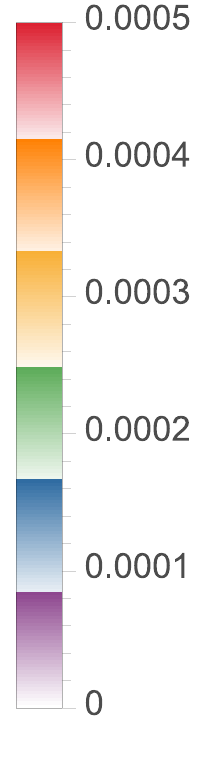}
	\caption{Energy current intensity $\pa{\I^E}^2$ (left) and angular momentum current intensity $\pa{\I^L}^2$ (right) evaluated at Poincar\'e time $t=0$ for the solution $F=\Re F_{(2,0)}$. Full animations are available at \cite{video}.}
	\label{Fig:Current}
	\end{center}
\end{figure}

To visualize the physical properties of these solutions, we animate the electric and magnetic field strengths $E^2=E_\nu E^\nu$ and $B^2=B_\nu B^\nu$ corresponding to the real solution $F=\Re F_{(2,0)}$, where
\begin{align}
	E_\nu\equiv-U^\mu F_{\mu\nu},\qquad B_\nu\equiv U^\mu(\star F)_{\mu\nu},
\end{align}
and $U^\mu=(1,0,0,0)$ is the 4-vector of a static observer in Poincar\'e coordinates. The result is accessible at \cite{video}. Figure \ref{Fig:Intro} contains screenshots of the simulations at $t=0$. Likewise, we also animated the energy and angular momentum currents $\I^E$ and $\I^L$. Figure \ref{Fig:Current} depicts screenshots of these simulations at $t=0$. In all these figures, the color scheme is in natural units.

\begin{figure}[h!]
	\begin{center}
	\href{http://www.youtube.com/playlist?list=PLsrfyTK-g7cP-_8F7A5Zb71K_94_gaXgn}
	{\includegraphics[width=0.4\textwidth]{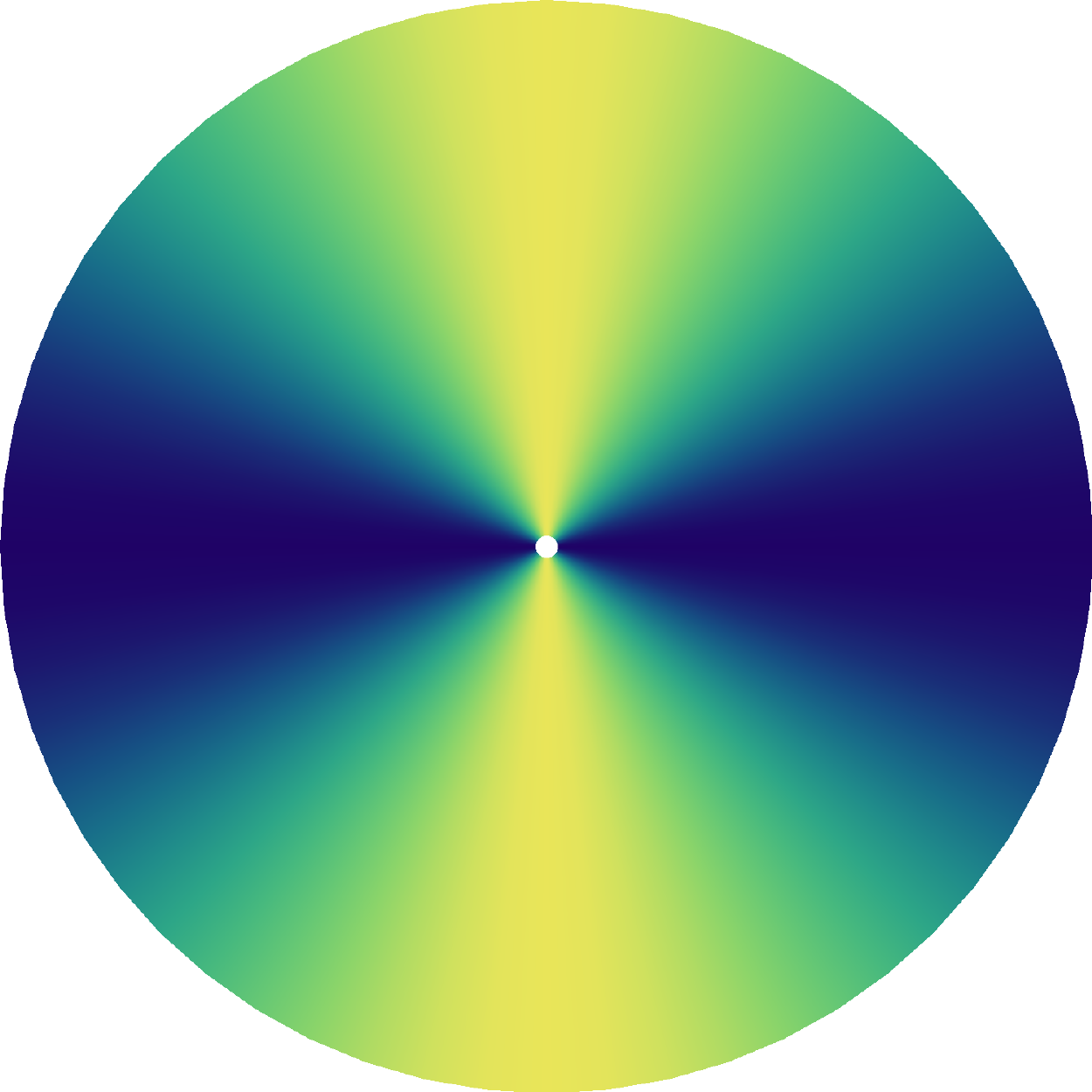}}\qquad
	\href{http://www.youtube.com/playlist?list=PLsrfyTK-g7cP-_8F7A5Zb71K_94_gaXgn}
	{\includegraphics[width=0.4\textwidth]{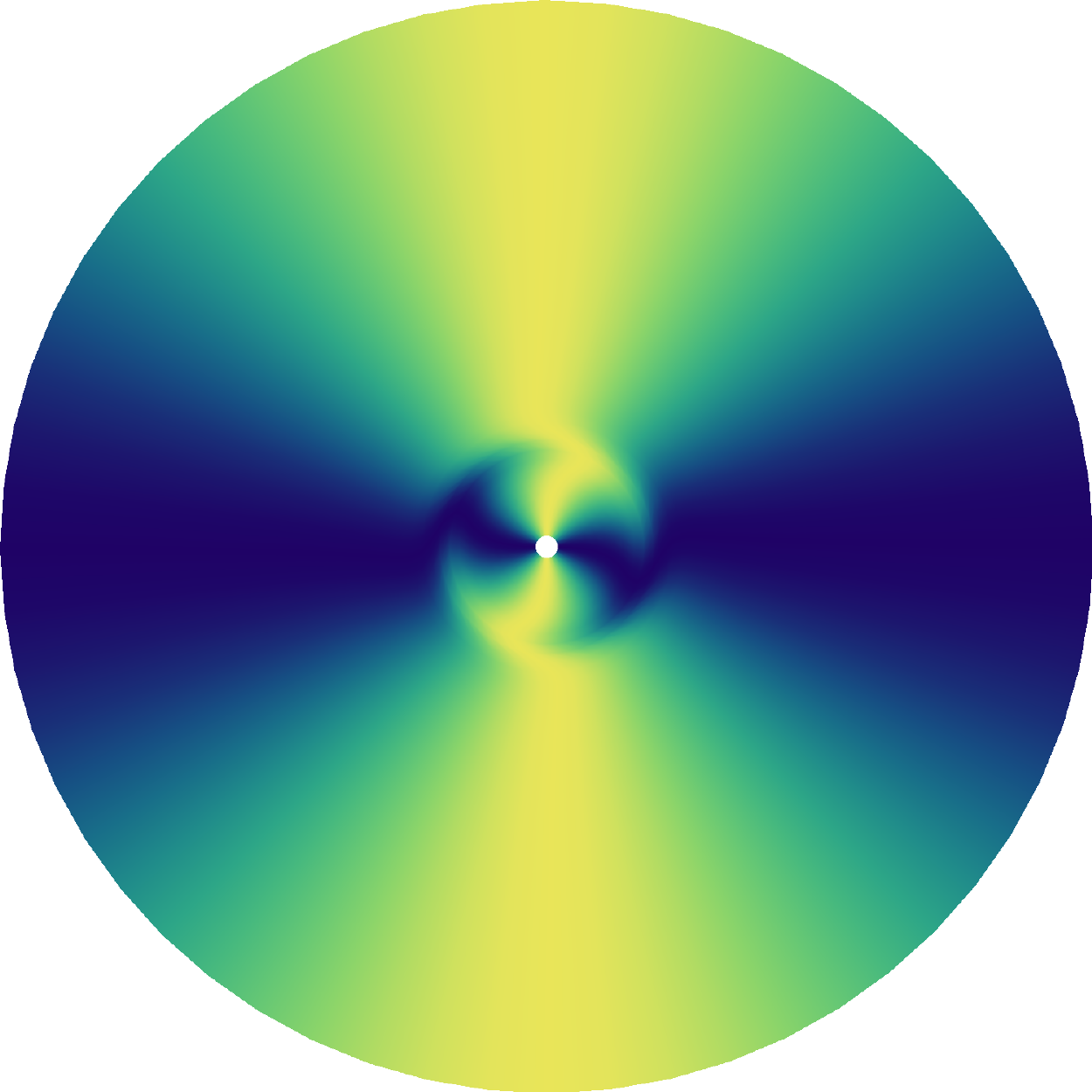}}
	\caption{Current $\J$ for the solution $\Re F_{(2,0)}$. At $t=0$ we begin to evolve the colored points on the left, resulting at $t=5$ in the image on the right. A full animation is accessible at \cite{video}.}
	\label{Fig:CurrentJ}
	\end{center}
\end{figure}

Finally, we also animated the flow of the current $\J$. Figure \ref{Fig:CurrentJ} depicts screenshots of the simulations at $t=0$ and at $t=5$.

%%%%%%%%%%%%%%%%%%% SUBSECTION %%%%%%%%%%%%%%%%%%%
\subsection{Energy and angular momentum flux}

\begin{figure}[h!]
	\begin{center}
	\includegraphics[width=0.35\textwidth]{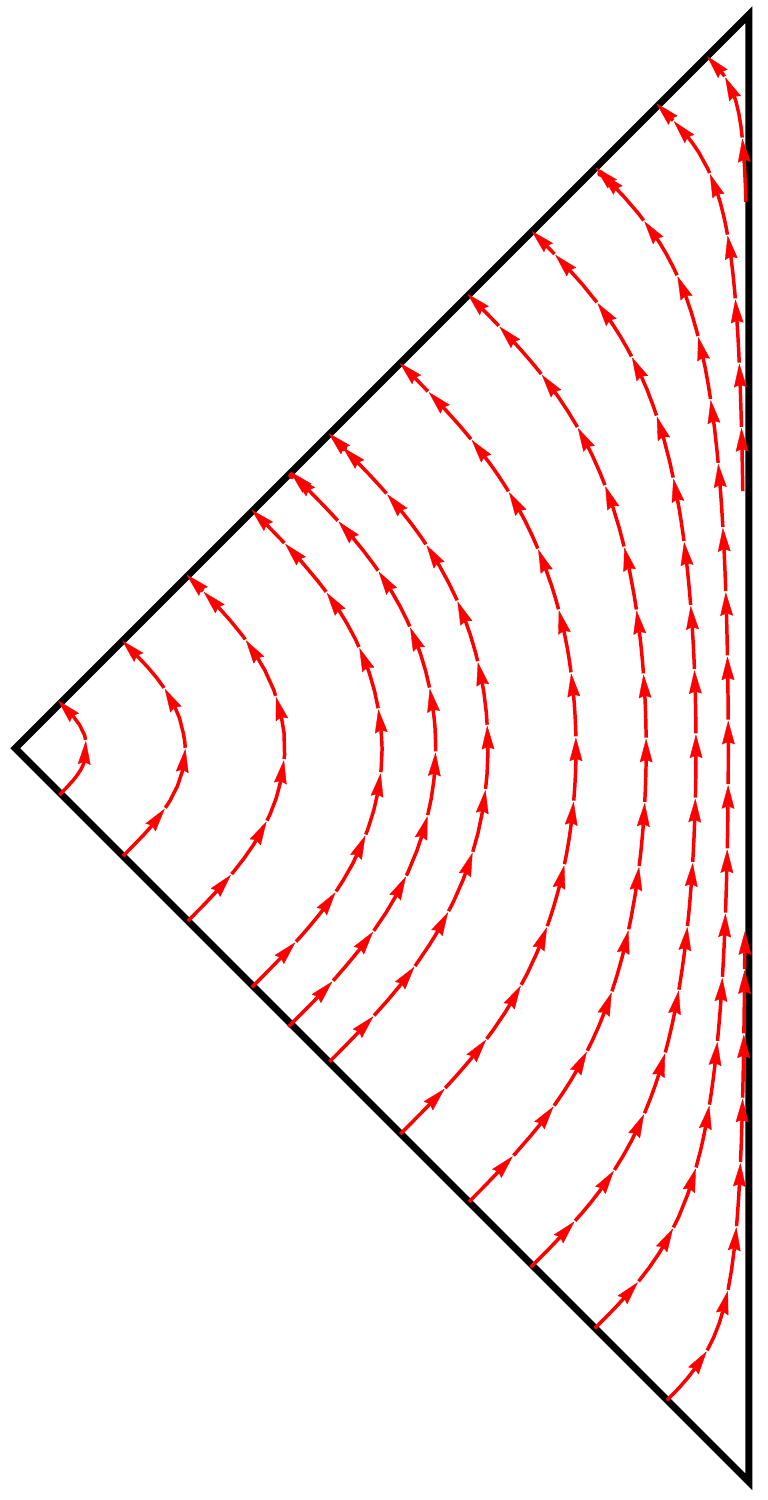}\qquad\qquad
	\includegraphics[width=0.35\textwidth]{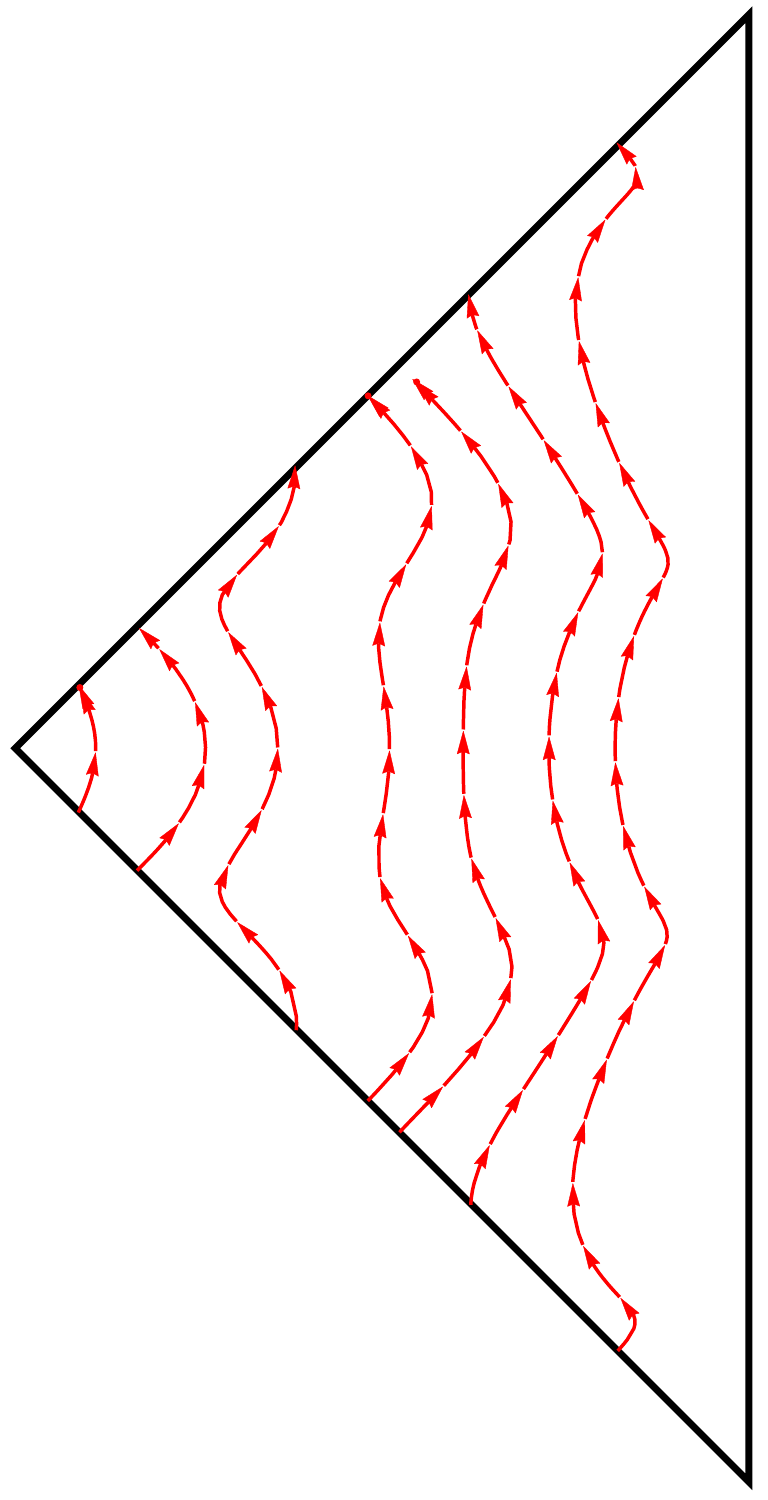}
	\caption{Diagram of the energy flux when $h\neq1$ and $\theta=0$ (left) or $\theta=\frac{\pi}{4}$ (right).}
	\label{Fig:Fluxes}
	\end{center}
\end{figure}

For the solutions $\Re F_{(h,0)}$, the energy and angular momentum fluxes \eqref{eq:fluxes} at the horizon are
\begin{align}
	\label{eq:RealFlux}
	\E_H^\pm=\pm2\sqrt{2}\Lambda\br{P_h'\pa{\sin{\pm\tau}}^{h+1}\cos{(h+1)\tau}}^2,\qquad
	\L_H^\pm=0.
\end{align}
Likewise, for the solutions $\Im F_{(h,0)}$,
\begin{align}
	\label{eq:ImaginaryFlux}
	\E_H^\pm=\pm2\sqrt{2}\Lambda\br{P_h'\pa{\sin{\pm\tau}}^{h+1}\sin{(h+1)\tau}}^2,\qquad
	\L_H^\pm=0.
\end{align}
In both cases, the fluxes out of the boundary of NHEK vanish for $h>\frac{1}{2}$:
\begin{align}
	\E_\infty=\L_\infty=0,
\end{align}
Plugging these expressions into \eqref{eq:totalfluxes} yields
\begin{align}
	\Delta E_H^+=-\Delta E_H^-,\qquad\Delta E_B=0,
\end{align}
which is of course consistent with \eqref{eq:fluxcondition}. This case is illustrated in Figure \ref{Fig:Fluxes}. On the other hand, when $h<\frac{1}{2}$, the energy flux density $\E_\infty$ at the boundary becomes divergent. For the boundary case $h=\frac{1}{2}$, it is nonzero but finite: for the solutions $\Re F_{(1/2,0)}$,
\begin{align}
	\E_\infty=2\sqrt{2}\,\Lambda\,{P_{1/2}'}^2(1+\cos\tau)\sin{\tau}\cos^2{\frac{\tau}{2}},
\end{align}
while for the solutions $\Im F_{(1/2,0)}$,
\begin{align}
	\E_\infty=-2\sqrt{2}\,\Lambda\,{P_{1/2}'}^2\cos\tau\sin{\tau}\cos^2{\frac{\tau}{2}}.
\end{align}
In either situation, the total flux through the boundary $\Delta E_B$ is still zero, which is consistent with the fact that the energy flux out of the future horizon equals that into the past horizon.

Finally, note from \eqref{eq:RealFlux} and \eqref{eq:ImaginaryFlux} that the total horizon fluxes $\Delta E_H^\pm$ are only finite when $h\ge-1$. For $h<-1$, even though these quantities diverge, the relation $\Delta E_H^+=\Delta E_H^-$ still holds.

%In spite of this equality, these solutions may nevertheless model energy extraction: matter incident on an %initially unexcited black hole could excite disturbances of the magnetosphere which are approximated by one %of the above solutions near the future (but not past) horizon. If and how this might occur is an interesting topic %for future studies.

%%%%%%%%%%%%%%%%%%%% SECTION %%%%%%%%%%%%%%%%%%%%
\section{Null $h=1$  solutions }
\label{sec:MariaSolution}

In section \ref{sec:Solutions}, we presented an infinite family of axisymmetric solutions to the force-free equation in NHEK given by highest-weight representations of $\SL(2,\R)$ with highest-weight $h\in\R-\cu{1}$. The special case $h=1$  degenerated to a trivial solution of the free Maxwell equations. In this section, we present a different highest-weight solution with $h=1$, which is nontrivial but has ``null" $F^2=0$. We suspect that it is some kind of limit of  the null solutions for full Kerr found in \cite{Brennan:2013jla}, but have not verified the details.

%%%%%%%%%%%%%%%%%%% SUBSECTION %%%%%%%%%%%%%%%%%%%
\subsection{Null highest-weight solution }
\label{subsec:HWreph1}

Consider the Ansatz
\begin{align}
	A_{(1,0)}=\Psi\Lambda\tilde{P}\ed\theta,
\end{align}
where in the last line we introduced a scalar function
\begin{align}
	\Psi(\tau,\psi)\equiv-e^{-i(\tau+\psi)},
\end{align}
while $\tilde{P}(\theta)$ can be an arbitrary regular function. $\Psi(\tau,\psi)$ is a $\U(1)\times\U(1)$ eigenfunction 
\begin{align}
	\L_{L_0}\Psi&=\Psi,\\
	\L_{Q_0}\Psi&=0.
\end{align}
However, it does not lie in a scalar highest-weight representation of $\SL(2,\R)$ because it is not annihilated by $L_+$. Instead, it obeys
\begin{align}
	\L_{L_+}\Psi=1.
\end{align}
Note also that is just a complex phase: $\Psi^*\Psi=1$. It follows that $A_{(1,0)}$ obeys
\begin{align}
	\L_{L_+}A_{(1,0)}&=\Psi^*A_{(1,0)}=\Lambda\tilde{P}\ed\theta,\\
	\label{L0invariant}
	\L_{L_0}A_{(1,0)}&=A_{(1,0)},\\
	\label{Q0invariant}
	\L_{Q_0}A_{(1,0)}&=0.
\end{align}
Observe that it is annihilated by $L_+$ up to a gauge transformation. The corresponding 2-form field strength $F_{(1,0)}\equiv\ed A_{(1,0)}$ is 
\begin{align}
	\label{NullSolution}
	F_{(1,0)}&=-i\Psi\Lambda\tilde{P}\ed\theta\wedge(\ed\tau+\ed\psi)\\
	&=\frac{i\Phi\Lambda\tilde{P}}{2J\Gamma}\ed\theta\wedge\pa{\Psi L_+-L_0-Q_0},
\end{align}
where $\Phi(\tau,\psi)$ is defined in \eqref{Phi}. From \eqref{NullSolution}, it is easily checked that $\L_{L_+}F_{(1,0)}=0$, which together with \eqref{L0invariant}--\eqref{Q0invariant} implies that $F_{(1,0)}$ is $U(1)$-invariant and forms a highest-weight representation of $\SL(2,\R)$ with highest weight $h=1$. The associated current $\J_{(1,0)}\equiv\ed^\dagger F_{(1,0)}$ is given by
\begin{align}
	\J_{(1,0)}&=\frac{i\Psi\pa{\Lambda\tilde{P}'+2\Lambda'\tilde{P}}}{2J\Gamma}
	\pa{\ed\tau+\ed\psi}\\
	&=\frac{i\Phi\pa{\Lambda\tilde{P}'+2\Lambda'\tilde{P}}}{(2J\Gamma)^2}
	\pa{\Psi L_+-L_0+Q_0}.
\end{align}
while
\begin{align}
	\label{HodgePrimaryh1}
	\star F_{(1,0)}&=-i\Psi\Lambda^2\tilde{P}
	\br{\cot{\psi}\ed\tau\wedge\ed\psi+\pa{\ed\tau+\ed\psi}\wedge\ed\varphi}\\
	&=\frac{i\Phi\tilde{P}}{(2J\Gamma)^2}Q_0\wedge\pa{\Psi L_+-L_0+Q_0}.
\end{align}
Since both are proportional to $\ed\tau+\ed\psi$ (or $\Psi L_+-L_0+Q_0$)
\begin{align}
	\J_{(1,0)}\wedge\star F_{(1,0)}=0
\end{align}
and  $F_{(1,0)}$ is a force-free solution. Note that it is null in the sense that
\begin{align}
	F_{(1,0)}^2=0.
\end{align}
The stress energy-momentum tensor of this solution takes the simple form
\begin{align}
	T_{(1,0)}^{\mu\nu}
	=2J\Gamma\pa{\frac{\Lambda\tilde{P}}{\Lambda\tilde{P}'+2\Lambda'\tilde{P}}}^2
	\J_{(1,0)}^\mu\J_{(1,0)}^\nu.
\end{align}
Hence we may interpret this solution as describing a pressureless perfect fluid.

%%%%%%%%%%%%%%%%%%% SUBSECTION %%%%%%%%%%%%%%%%%%%
\subsection{Descendants, reality conditions and general solutions}

Let us now consider descendants. The situation here is similar to the generic case of section \ref{subsec:Descendants}. Using the relation
\begin{align}
	\L_{L_-}(\ed\tau+\ed\psi)=\Psi\pa{\ed\tau+\ed\psi},
\end{align}
it is easily seen that all descendants of both $ \J_{(1,0)}$ and $\star F_{(1,0)}$ (as well as real or imaginary parts thereof) are proportional to $\ed\tau+\ed\psi$. Hence any linear combination of the real or imaginary parts of any descendants of $ A_{(1,0)}$ is a force-free solution.

%%%%%%%%%%%%%%%%%% SUBSUBSECTION %%%%%%%%%%%%%%%%%%
\subsubsection{Energy and angular momentum flux}

\begin{figure}[h!]
	\begin{center}
	\includegraphics[width=0.35\textwidth]{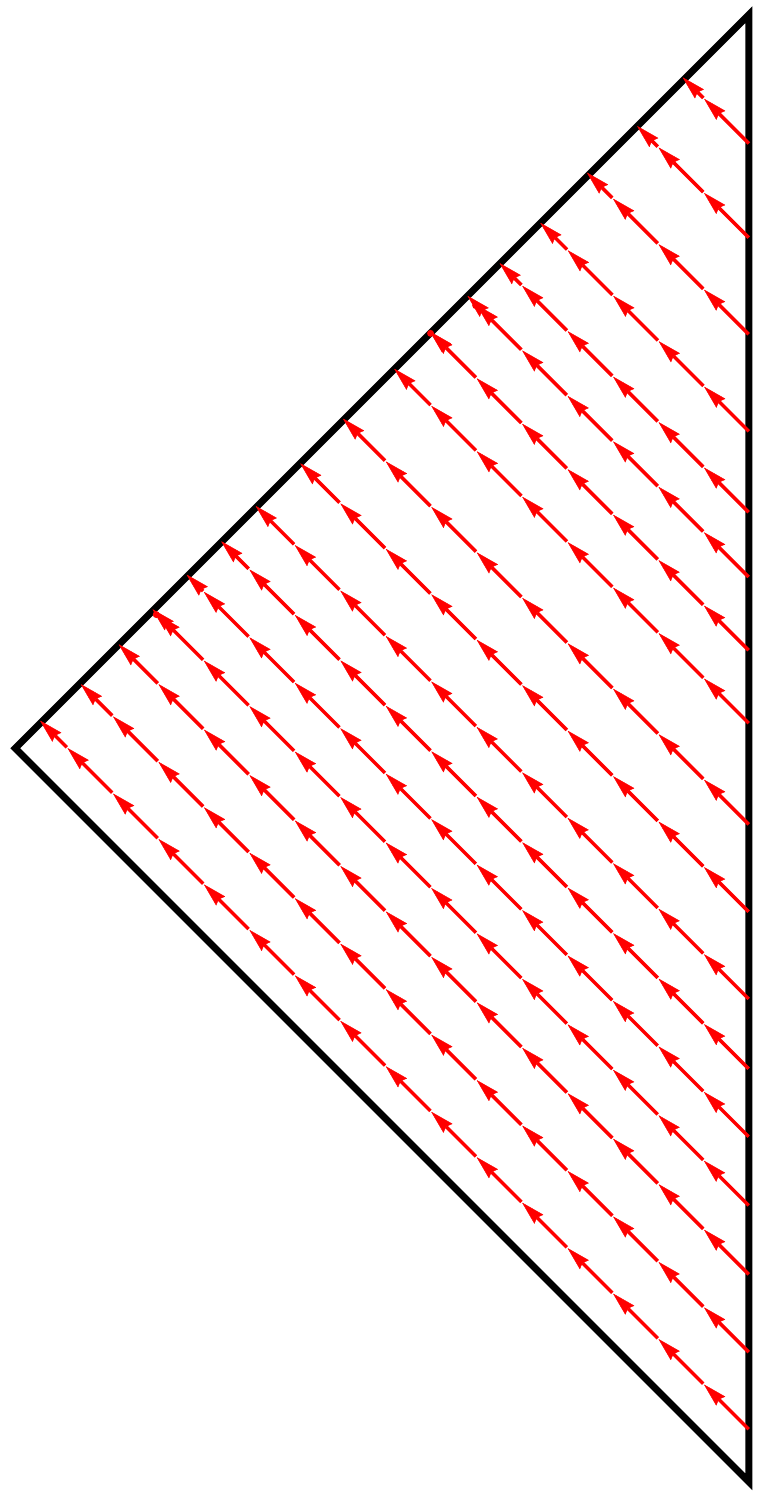}
	\caption{Diagram of the energy flux when $h=1$.}
	\label{Fig:Fluxesh1}
	\end{center}
\end{figure}

For the solution $\Re F_{(1,0)}$, the energy and angular momentum fluxes at the horizon \eqref{eq:fluxes} are
\begin{align}
	\E_H^+=8\sqrt{2}\Lambda^3\pa{\tilde{P}\sin^2{\tau}\sin{2\tau}}^2,\qquad
	\E_H^-=0,\qquad
	\L_H^\pm=0.
\end{align}
The energy flux at the boundary is
\begin{align}
	\E_\infty=\sqrt{2}\Lambda^3\br{\tilde{P}\pa{1+\cos{\tau}}\sin{\tau}}^2.
\end{align}
Likewise, for the solution $\Im F_{(1,0)}$,
\begin{align}
	\E_H^+=8\sqrt{2}\Lambda^3\pa{\tilde{P}\sin^2{\tau}\cos{2\tau}}^2,\qquad
	\E_H^-=0,\qquad
	\L_H^\pm=0,
\end{align}
and
\begin{align}
	\E_\infty=\sqrt{2}\Lambda^3\br{\tilde{P}\pa{1+\cos{\tau}}\cos{\tau}}^2.
\end{align}
As for the angular momentum fluxes at the boundary of the throat, they vanish in both cases
\begin{align}
	\L_\infty=0.
\end{align}
Plugging these expressions into \eqref{eq:totalfluxes} yields
\begin{align}
	\Delta E_H^+=-\Delta E_B,\qquad\Delta E_H^-=0,
\end{align}
which is still consistent with \eqref{eq:fluxcondition}. This solution is illustrated in Figure \ref{Fig:Fluxesh1}.

%%%%%%%%%%%%%%%%% ACKNOWLEDGMENTS %%%%%%%%%%%%%%%%%
\section*{Acknowledgements}
We are grateful to Gim-Seng Ng for useful conversations, and to Samuel Gralla and Achilleas Porfyriadis for their helpful comments. We also thank Christopher Wolfram for his assistance with the creation of video animations in Mathematica. This work was supported in part by NSF grant 1205550 and the Fundamental Laws Initiative at Harvard. M.J.R. is supported by the European Commission - Marie Curie grant PIOF-GA 2010-275082.

\appendix

%%%%%%%%%%%%%%%%%%%% APPENDIX %%%%%%%%%%%%%%%%%%%%
\section{Fluxes in Poincar\'{e} and global coordinates}
\label{appendix:Details}

Here we elaborate some subtleties that arise in the computations of the energy and angular momentum fluxes for the NHEK geometry in Poincar\'{e} or global coordinates.

To determine the unit vector normal to the spherical shell of constant radius $r=r_0$, note that in Poincar\'{e} coordinates (\ref{eq:poincareNHEK}) it is defined by the vanishing of
\begin{align}
	f(t,r,\theta,\phi)\equiv r-r_0.
\end{align}
We may thus obtain a normal vector $\xi$ to this hypersurface by defining
\begin{align}
	\xi\equiv\nabla^\mu f\pd_\mu=\frac{r^2}{2J\Gamma}\pd_r.
\end{align}
As expected, this vector field is spacelike for $r>0$,
\begin{align}
	\xi\cdot\xi=\frac{r^2}{2J\Gamma},
\end{align}
but it becomes null at $r=r_H=0$, the location of the event horizon. In fact, it vanishes identically there, $\xi|_{r=r_H}=0$, and we should replace it by $H_+$, which is the null vector normal to the horizon. This complication is an artifact of the coordinate system, and does not arise in the global coordinates \eqref{eq:globalNHEK}, which are smooth across the horizon. Away from the horizon, we may normalize $\xi$ to obtain a unit normal vector
\begin{align}
	\label{NormalVectorPoincare}
	n\equiv\frac{\xi}{\sqrt{\xi\cdot\xi}}=\frac{r}{\sqrt{2J\Gamma}}\pd_r.
\end{align}
The flux densities of energy and angular momentum at the boundary of the throat ($r\to\infty$ in Poincar\'{e} coordinates) are then
\begin{align}
	\E_{\infty}&\equiv\lim_{r_0\to\infty}\sqrt{-\sigma}\,n^\nu\I_\nu^E,\\
	\L_{\infty}&\equiv\lim_{r_0\to\infty}\sqrt{-\sigma}\,n^\nu\I_\nu^L.
\end{align}
The induced 3-metric on the hypersurface $r=r_0$ has determinant $\sqrt{-\sigma}=(2J\Gamma)^{3/2}\Lambda r_0$, which grows linearly in the radius.

For the special case of the event horizon at $r=r_H=0$, where $n$ diverges (since $\xi\cdot\xi=0$), we must instead define the energy and angular momentum flux densities threading the black hole, denoted by $\E_H^\pm$ and $\L_H^\pm$ respectively, by
\begin{align}
	\E_H^\pm&=\sqrt{\gamma}\pa{\pm\sqrt{2}H_+^\nu}\I_\nu^E\Big|_{r=r_H},\\
	\L_H^\pm&=\sqrt{\gamma}\pa{\pm\sqrt{2}H_+^\nu}\I_\nu^L\Big|_{r=r_H}.
\end{align}
The factor of $\sqrt{2}$ introduced here compensates for the one in the definition of $H_+$. The determinant of the induced metric on the 2-sphere is $\sqrt{\gamma}=2J\Gamma\Lambda$, which is independent of radius. These are the rate of energy and angular momentum extraction from the black hole per unit solid angle. Great care must be taken when evaluating these quantities in Poincar\'{e} coordinates: noting from \eqref{eq:poincare2global} that $t\propto r^{-1}$, we see that as we send $r\to 0$, we must simultaneously push $t\to\infty$. Therefore, we can reach the (future or past) horizon by simply defining $r=\epsilon$, $t=t_0\pm1/\epsilon$ and taking the limit $\epsilon\to0$.

These quantities are more easily computed in global coordinates (\ref{eq:globalNHEK}), which are smooth across the horizon. By \eqref{eq:poincare2global}, the event horizon at $r=r_H=0$ becomes the hypersurface
\begin{align}
	\label{eq:Globalhorizon}
	\tau=\pm\psi,
\end{align}
where the sign depends on whether one is at the future or past horizon of the Poincar\'{e} patch. Also, the boundary of the throat ($r\rightarrow\infty$ in Poincar\'{e} coordinates) becomes the hypersurface $\psi=\pi$ in global coordinates. This explains \eqref{eq:fluxes}. As a consistency check, note that in global coordinates,
\begin{align}
	f(\tau,\psi,\theta,\varphi)&\equiv\frac{\cos{\tau}-\cos{\psi}}{\sin{\psi}}-r_0,\\
	\Longrightarrow\qquad\xi&\equiv\nabla^\mu f\pd_\mu
	=\frac{1}{2J\Gamma}\br{\sin{\tau}\sin{\psi}\pd_\tau
		+\pa{1-\cos{\tau}\cos{\psi}}\pd_\psi+\sin{\tau}\cos{\psi}\pd_\varphi}.
\end{align}
As expected, this vector field is spacelike,
\begin{align}
	\xi\cdot\xi=\frac{1}{2J\Gamma}\pa{\frac{\cos{\tau}-\cos{\psi}}{\sin{\psi}}}^2>0,
\end{align}
except at the event horizon $\tau=\pm\psi$, where it becomes null and agrees (up to a normalization factor) with 
\begin{align}
	\pm\frac{H_+}{\sqrt{2}}\,\bigg|_{\tau=\pm\psi}=2J\Gamma\xi\,\big|_{\tau=\pm\psi}.
\end{align}
Away from the horizon, we may normalize $\xi$ to obtain a unit normal vector
\begin{align}
	n\equiv\frac{\xi}{\sqrt{\xi\cdot\xi}}
	=\frac{\sin{\tau}\sin{\psi}\pd_\tau+\pa{1-\cos{\tau}\cos{\psi}}\pd_\psi+\sin{\tau}\cos{\psi}\pd_\varphi}
	{\sqrt{2J\Gamma}\pa{\cos{\tau}-\cos{\psi}}\csc{\psi}},
\end{align}
which agrees with the expression \eqref{NormalVectorPoincare} for $n$ in Poincar\'{e} coordinates.

%%%%%%%%%%%%%%%%%%%% APPENDIX %%%%%%%%%%%%%%%%%%%%
\section{Analysis of $P_h$}
\label{appendix:Analysis}

In order to determine the $\theta$-dependence of the highest-weight solutions, we need to determine the behavior of the function $P_h(\theta)$ and solve \eqref{thetaode}
\begin{align}
	\pd_\theta\pa{\Lambda\pd_\theta P_h}+h(h-1)\Lambda P_h=0.
\end{align}
By performing a suitable coordinate transformation to a new variable $z=\sin^2\theta$, we may put this equation into the form of a generalized Heun equation, namely
\begin{align}
	P_h''(z)+\pa{\frac{\gamma}{z}+\frac{\delta}{z-1}+\frac{\epsilon}{z-a}}P_h'(z)
	+\frac{\alpha\beta\pa{z-q}}{z(z-1)(z-a)}P_h(z)=0,
\end{align}
where $\alpha+\beta+1=\gamma+\delta+\epsilon$ and $q$ is an accessory parameter. In our case, these parameters are given by $\gamma=1$, $\delta=1/2$, $\epsilon=-1$, $\alpha=-h/2$, $\beta=(h-1)/2$ and $q=2$. The 4 regular singular points of this equation are located at $z=z_0$, with $z_0\in\cu{0,1,a=2,\infty}$. The corresponding roots $(t_1,t_2)$ of the indicial equation are $(0,1-\gamma)$, $(0,1-\delta)$, $(0,1-\epsilon)$ and $(\alpha,\beta)$, respectively.

We may solve \eqref{thetaode} by expanding in a power series around each singularity $z_0$ (Frobenius' method). Since $z=\sin^2\theta$, the interval of interest to us, $\theta\in\br{0,\pi/2}$, gets mapped to the interval $z\in\br{0,1}$; the poles $\theta=0,\pi$ are mapped to $z=0$, while the $\theta=\pi/2$ plane is mapped to $z=1$. It therefore suffices for us to use the power series solutions \eqref{heunpowerseries} and these will converge everywhere on $\theta\in\br{0,\pi}$.

Depending on the nature of the roots of the indicial equation, there are three forms for the two linearly independent solutions on the intervals. On the one hand, at $\theta=\pi/2$ (or correspondingly, the singularity $z_0=1$), the roots of the indicial equation are $(t_1,t_2)=(0,1-\delta)$, hence $t_1-t_2\ne N,\ N\in\mathbb{Z}$ and therefore we can obtain the two independent solutions to the equation from the power series
\begin{align}
	\label{heunpowerseries}
	P_h^{(1)}(z)=z^{t_1}\sum_{n=0}^\infty \tilde{a}^{(1)}_n\,(z-z_0)^n
	\qquad\text{and}\qquad
	P_h^{(2)}(z)=z^{t_2}\sum_{n=0}^\infty \tilde{a}^{(2)}_n\,(z-z_0)^n.
\end{align}
These have a radius of convergence of $\ab{z-z_0}<r_0$, where $r_0$ corresponds to the distance to the closest singularity. Only one of the two solutions is symmetric about the $\theta=\pi/2$ plane. On the other hand around $z_0=0$ the two roots of the indicial equation are $(t_1,t_2)=(0,1-\gamma)$ but since $\gamma=1$, the roots are repeated: $t_1=t_2=0$. Hence the solutions to the ODE are 
\begin{align}
	\label{heunpowerseries}
	P_h^{(1)}(z)=\sum_{n=0}^\infty a^{(1)}_n\,z^n
	\qquad\text{and}\qquad
	P_h^{(2)}(z)=P_h^{(1)}(z)\log(z)+\sum_{n=1}^\infty a^{(2)}_n\,z^n.
\end{align}

%%%%%%%%%%%%%%%%%%% BIBLIOGRAPHY %%%%%%%%%%%%%%%%%%%

\end{document}